\documentclass[useAMS,usenatbib]{mn2e}

\usepackage{amsmath,array,amsfonts}
\usepackage{amssymb}
\usepackage{graphicx}
\usepackage{placeins}
\usepackage{float}
\usepackage{natbib}
\usepackage{verbatim}
\usepackage{pstricks}

\title[Intrinsic size correlations in weak lensing]{Intrinsic size correlations in weak lensing}
\author[S. Ciarlariello et al.]{Sandro Ciarlariello$^1$\thanks{e-mail: sandro.ciarlariello@port.ac.uk}, Robert Crittenden$^1$, Francesco Pace$^{2,1}$ \\
$^1$ Institute of Cosmology and Gravitation, University of Portsmouth, Dennis Sciama Building, Portsmouth PO1 3FX, UK  \\
$^2$ Jodrell Bank Centre for Astrophysics, School of Physics and Astronomy, The University of Manchester, Manchester, M13 9PL, UK}

\begin{document}

\date{Accepted ? ;  Received ? ; in original form ?}

\pagerange{\pageref{firstpage}--\pageref{lastpage}} \pubyear{}

\maketitle

\label{firstpage}

\begin{abstract}
We present a simple model for describing intrinsic correlations for galaxy sizes based on the halo model. Studying these correlations is important both to improve our understanding of galaxy properties and because they are a potential systematic for weak lensing size magnification measurements.  Our model assumes that the density field drives these intrinsic correlations and we also model the distribution of satellite galaxies.  We calculate the possible contamination to measurements of lensing convergence power spectrum from galaxy sizes, and show that the cross-correlation of intrinsic sizes with convergence is potentially an important systematic.  We also explore how these intrinsic size correlations may affect surveys with different redshift depth. We find that, in this simple approach, intrinsic size correlations cannot be neglected in order to estimate lensing convergence power spectrum for constraining cosmological parameters.
\end{abstract}

\begin{keywords}
gravitational lensing: weak - methods: analytical, statistical
\end{keywords}

\section{Introduction}
Weak gravitational lensing has the potential to be one of the most powerful probes in cosmology, constraining models with great precision. 
By measuring small but coherent changes in the shape, brightness and size of background galaxies, weak lensing can tell us both about the background cosmology and the distribution of dark matter in the Universe.  
Weak lensing measurements thus far have focused primarily on the shape distortions, or shear; cosmic shear correlations were first detected in 2000 by several groups \citep{Bacon2000, Kaiser2000, Waerbeke2000, Wittman2000} and have since been significantly improved using surveys such as CFHTLens \citep{Heymans2012}.  Shear measurements of weak lensing are a critical component of future surveys such as Euclid\footnote{http://www.euclid-ec.org/} and LSST\footnote{http://www.lsst.org/lsst}.

Cosmic magnification, leading to coherent size and brightness distortions, has been also been observed but not to the same extent as shear. 
Magnification has been primarily probed through the cross-correlation between foreground galaxies and background objects selected with by their flux, known as flux magnification or magnification bias.  This was first detected using background quasars by \cite{Scranton2005} and other background sources, such as Lyman-break galaxies, have since been used to study the mass profiles of dark matter haloes \citep{Hildebrandt2009, Waerbeke2010b, Hildebrandt2011, Ford2012, Hildebrandt2013, Bauer2013}. 
Such galaxy-galaxy lensing can be combined with shear measurements as a complementary weak lensing probe which allow us to control systematics and cosmological parameters constraints \citep{Waerbeke2010, Duncan2014}.    

Cosmic magnification can also be detected directly using galaxy sizes and magnitudes \citep{Bartelmann1996} because size information is already available from a lensing survey, and this has recently been observed by \cite{Schmidt2011}.  \cite{Huff2014} also detected magnification using sizes measured by means of the Fundamental Plane relation for early-type galaxies. 
Following these measurements \cite{Casaponsa2013} studied the extent size magnification can be used as a complement to cosmic shear, investigating how observational limitations can affect this kind of measurement.  \cite{Heavens2013} showed that combining size and shape information from weak lensing measurements could, in principle, improve our current constraints on cosmological parameters obtained by means of only cosmic shear.  Recently, \cite{Alsing2014} extended this analysis to quantify the convergence dispersion expected from size measurements and the possible impact of intrinsic size correlations.  

Intrinsic correlations between the observed shapes and sizes of galaxies can arise via other physical mechanisms and mimic the effects of weak lensing.     
For cosmic shear, it has long been recognised \citep{Heavens2000, Catelan2000, Croft2000, Crittenden2001, Hirata2004} that intrinsic alignments of galaxies are important systematics and could lead to biases if not accounted for.  These intrinsic ellipticity correlations have been measured in several works \citep{Brown2002, Mandelbaum2006, Hirata2007, Faltenbacher2009, Okumura2009, Joachimi2010a,  Mandelbaum2011, Joachimi2011, Joachimi2013, Chisari2014, Sifon2014, Singh2014}.   The physical mechanisms for these 
correlations are not fully understood, and the mechanisms appear to depend on the galaxy type. The shape of elliptical galaxies is thought to reflect mainly the shape of the halo in which the galaxy is embedded and the halo shape is given by the gravitational tidal field on large scales. For disk galaxies, alignments can arise from angular momentum correlations. Indeed, if two disks spin along the same direction then they are seen under the same inclination by an observer. 

A number of methods have been proposed to mitigate these intrinsic alignments, either by removing pairs physically close \citep{King2002a, King2002b}, developing a model based on the halo model \citep{Schneider2010}, using a technique for boosting or nulling this intrinsic signal \citep{Joachimi2010b} or adopting path integral marginalisation over all the possible intrinsic alignment models \citep{Kitching2010}.  Without some attempt to correct for this systematic, very significant biases will appear in constraints from future measurements \citep{Bridle2008, Kirk2010, Kirk2012}. 

Here we attempt a similar investigation into whether there could be intrinsic correlations in the sizes of galaxies that would similarly bias the interpretation of magnification data.  From an observational point of view, the situation is unclear. There are some recent claims of dependence of galaxy size on the environment \citep{Cooper2012} as well as other claims where either no dependence has been found \citep{Rettura2010} or a possible anti-correlation has been found between environment and galaxy size \citep{Maltby2010}.  To estimate their impact on weak lensing, \cite{Alsing2014} modelled size correlations in a heuristic way; our aim here is to find a more physical model for these correlations. 

We investigate the degree to which intrinsic size correlations arise in a simple halo model, assuming the observed galaxy sizes correlate closely with the mass of the haloes and sub-haloes \citep{Kravtsov2013}.   Larger and more massive galaxies live in more massive haloes, and even if the sub-halo population is largely independent of the halo mass, the sizes of the largest sub-haloes will still be limited by the total halo mass.    We use this simple model to predict what would be observed for a magnification estimator based solely on the galaxy sizes, and how the intrinsic signal correlates with the true lensing convergence.   

The paper is organised as follows. In section \ref{lensing-sec} we introduce the lensing formalism and theory. In section \ref{halomodel} we discuss the halo model formalism and in section \ref{sizecorr} we apply it to intrinsic size correlations. In section \ref{2pointStat} the power spectra of intrinsic size correlations are calculated. In section \ref{results} results are shown and analysed and then we conclude in section \ref{conclusions}.

\section{Lensing magnification}
\label{lensing-sec}
Weak gravitational lensing by large scale structure can be observed both through shape distortion correlations (cosmic shear) and the magnification of distant galaxies. These two effects are described by the transformation matrix which maps the true galaxy source positions to their observed position on the sky,
\begin{equation}
A(\bmath\theta) = \left(\delta_{ij} - \frac{\partial^{2}\psi(\bmath\theta)}{\partial\theta_{i}\partial\theta_{j}} \right) =  \left(\begin{array}{cc} 1-\kappa-\gamma_{1} & -\gamma_{2} \\ -\gamma_{2} & 1-\kappa+\gamma_{1} \end{array} \right)\;,
\label{lensmatrix}
\end{equation}
where $\psi(\bmath\theta)$ is the two-dimensional gravitational potential, $\gamma_{1}= \frac{1}{2}(\psi_{,11}-\psi_{,22})$, $\gamma_{2} = \psi_{,12}$ where the comma in $\psi_{,i}$ represents the partial derivative of the gravitational potential with respect to the variable $\theta_{i}$ and $\kappa$ is the convergence. Indicating the cosmic shear by $\gamma$, we have: $\gamma = \gamma_{1} +\rm{i}\gamma_{2}$.
The determinant of this matrix gives the cosmic magnification $\mu$ of a surface area element:
\begin{equation}
\mu = \frac{1}{\det{A}}=[(1-\kappa)^2-|\gamma|^2]^{-1}\;.
\label{mag}
\end{equation}
In the weak lensing regime $|\kappa|$ and $|\gamma|\ll 1$, so the magnification is approximately $\mu \simeq 1 + 2\kappa$.

Given eq. (\ref{mag}) for the relation between magnified and intrinsic surface area element, we can derive the relation between magnified and intrinsic angular sizes.  The angular size, $\lambda$, of an object becomes 
\begin{equation}
\lambda_{\rm{O}}=(1+\kappa)\lambda_{\rm{I}}, 
\end{equation}
where the subscripts stand for the observed $(O)$ and intrinsic $(I)$  angular size of the galaxy; we define the intrinsic angular size to be the square root of the solid angle of the galaxy image. As pointed out by \cite{Heavens2013}, this definition for the galaxy size is uncorrelated with shear.
In the weak lensing limit, this can be written 
\begin{equation}
\ln{\frac{\lambda_{\rm{O}}}{\bar{\lambda}}} \simeq \kappa + \ln{\frac{\lambda_{\rm{I}}}{\bar{\lambda}}}\;,
\end{equation}
where $\bar{\lambda}$ is the mean angular size at a given redshift.
Then we use as our estimator the following one \citep{Schmidt2011, Heavens2013, Bacon2014}:
\begin{equation}
\hat{\kappa} = \ln{\frac{\lambda_{\rm{O}}}{\bar{\lambda}}} - \left\langle \ln{\frac{\lambda_{\rm{O}}}{\bar{\lambda}}} \right\rangle\;,
\end{equation}
which has zero mean.
Note that, relative to their average at a given redshift,  the physical size of a galaxy $r$ is essentially a proxy for its observed angular size $\lambda$ because the angular diameter distance $D_{\rm{A}}(z)$ is the same for the average:  
\begin{equation}
\frac{r(z)}{\bar{r}(z)} = \frac{\lambda D_{\rm{A}}(z)}{\bar{\lambda} D_{\rm{A}}(z)} = \frac{\lambda}{\bar{\lambda}}\;.
\label{angtophys}
\end{equation}

For any given galaxy, its observed size will be determined more by its intrinsic size than by its magnification, so any individual measurement will be dominated by this intrinsic size dispersion.  But by averaging many such measurements over a patch where the magnification is coherent, one can reach a regime where the magnification dominates.  
However, this assumes that the average intrinsic sizes are uncorrelated; if there are intrinsic correlations in sizes, so that 
$\langle r \rangle_{\rm patch} \ne \bar{r} $ then this could be wrongly interpreted as magnification. 
The magnification estimator will effectively have two contributions, the true convergence and the intrinsic contribution: 
\begin{equation}
\hat{\kappa} = \kappa + \kappa_{\rm{I}}\;.
\end{equation}
Here, $\kappa_{\rm{I}}$ is the contribution to the size magnification estimator arising from the intrinsic sizes; in particular,
\begin{equation}
\kappa_{\rm{I}} \equiv  \ln{\frac{\lambda_{\rm{I}}}{\bar{\lambda}}} - \left\langle \ln{\frac{\lambda_{\rm{I}}}{\bar{\lambda}}} \right\rangle\;.
\end{equation}
The primary observables are the two point moments of the estimator, which has three contributions; in Fourier space, these are written as 
\begin{equation}
C_{\hat{\kappa}}(\ell) = C_{\kappa}(\ell) + 2C_{\kappa\kappa_{\rm{I}}}(\ell) + C_{\kappa_{\rm{I}}}(\ell)\;.
\end{equation}
The lensing auto-correlation is well understood, and here we investigate the other terms in a simple halo model. 

The lensing convergence power spectrum, $C_\kappa$, is given by means of the Limber approximation \citep{Limber1954}:
 \begin{equation}
 C_\kappa(\ell) = \int_0^{\chi_{\rm{hor}}} d\chi \, 
\frac{q^{2}(\chi)}{[f_{K}(\chi)]^2} \, P_\delta \left( \frac{\ell}{f_{K}(\chi)},\chi \right)\;,
\label{convergence}
\end{equation}
where $P_{\delta}$ is the matter power spectrum, $\chi$ is the comoving distance along the line of sight, $\chi_{\rm{hor}}$ is the comoving horizon distance and $f_{K}(\chi)$ is the comoving angular diameter distance.  The weighting function 
\begin{equation}
q(\chi) = \frac{3 H_0^2 \Omega_{\rm{m,0}}}{2c^2} \frac{f_{K}(\chi)}{a(\chi)}\int_\chi^{\chi_{\rm{hor}}}\, d\chi'\ n(\chi') 
\frac{f_{K}(\chi'-\chi)}{f_{K}(\chi')}\;,
\label{lensingweight}
\end{equation}
where $a$ is the dimensionless scale factor,  $c$ is the speed of light, $H_0$ is the Hubble constant, $\Omega_{\rm{m,0}}$ is the present matter density parameter and $n(\chi)d\chi$ is the effective number of galaxies in $d\chi$, normalized so that $\int n(\chi)d\chi = 1$.
The radial function $f_{K}(\chi)$ depends on $K$, the inverse square of curvature radius in units of $H_{0}/c$,  as follows:
\begin{equation}
 f_{K}(\chi)=\left\{
      \begin{array}{ll}
        \sqrt{K}\sin(\sqrt{K}\chi) & \quad K>0 \\
      \chi & \quad K=0 \;, \\
      \sqrt{-K}\sinh(\sqrt{-K}\chi) & \quad K<0 .
      \end{array}
    \right.
\end{equation}
For simplicity, below we will assume $K=0$.

\section{The halo model }
\label{halomodel}
\subsection{Overview}
\label{overview}
Here we describe a simple model for how galaxy sizes may be intrinsically correlated, based on the halo model formalism \citep{Scherrer1991, Seljak2000, Cooray2002, ShethJain2002}.  We first describe our implementation of the halo model itself, and will discuss its implications for galaxy sizes in the next section. 

The halo model assumes the mass in the Universe is distributed into distinct haloes, whose large scale distribution is described by mass dependent two-point (and potentially higher order) correlations.   
A central galaxy is associated with the halo centre, and satellite galaxies are distributed around it with some profile probability density.     
The satellites are associated with sub-haloes, which have a distribution of mass which in principle depends on the mass of the halo in which they sit.   

A complete specification of the halo model requires knowing the halo mass function and the distribution of sub-halo masses within a halo; it also requires knowing the probability density profile of how sub-haloes are distributed in a halo and understanding the statistics of how haloes are distributed on large scales, usually parameterised by the mass dependent bias function.  

\subsection{Elements of the halo model}

Throughout we will indicate halo masses and sizes with $M, R$ and sub-halo (or satellite) masses and sizes with $m, r$.

\subsubsection{Mass function} 
The comoving number density of collapsed haloes with mass between $M$ and $M+dM$ is described by the halo mass function $n(M, z)$
\begin{equation}
\frac{M^{2}}{\bar{\rho}}n(M)\frac{dM}{M} = \nu f(\nu)\frac{d\nu}{\nu}\;; 
\end{equation}
here $\bar{\rho}$ is the comoving background density and $\nu \equiv \delta_{c}^{2}(z)/\sigma^{2}(M,z)$ is the ratio of the critical density for the spherical collapse (squared) to the variance of a halo with mass $M$. 
The function $f(\nu)$ can be written as \citep{ShethTormen1999}
\begin{equation}
\nu f(\nu) = A(p)[1+(q\nu)^{-p}]\left(\frac{q\nu}{2\pi}\right)^{1/2}\exp{\left(-\frac{q\nu}{2}\right)}\;.
\end{equation}
\cite{Weinberg2003} found that, for $0.1 \leq \Omega_{m} \leq 1$ and $-1 \leq w \leq -0.3$, the critical density at a given redshift is accurately given by the following fitting function:
\begin{equation} 
\delta_{\rm c}(z) = \frac{3}{20}(12\pi)^{2/3}(1+\alpha\log_{10}{\Omega_{\rm{m}}(z)})\;,
\end{equation}
where $\alpha(w)$ is a function of the dark energy equation of state parameter $w$: 
\begin{equation}
\alpha(w) = 0.353w^{4} + 1.044w^{3} + 1.128w^{2} + 0.555w +  0.131\;.
\end{equation}
Here we will assume a cosmological constant ($w=-1$) for which $\alpha = 0.013$. 
Assuming spherical collapse, $p=0$, $A(p)=0.5$,  and $q=1$ \citep{PS1974}; alternatively, ellipsoidal collapse results in the Sheth-Tormen mass function where $p\simeq0.3$, $A(p)\simeq0.3222$, and $q\simeq0.75$ \citep{ShethTormen1999}.   In this paper we use the Sheth-Tormen formulation of the mass function because it provides better agreement with N-body simulations.

\subsubsection{Sub-halo mass function} 
For clustering statistics, it is sufficient to simply know how many galaxies are populating a halo of a given mass, known as the Halo Occupation Distribution (HOD).  However, for our purposes we also need to quantify the physical properties of satellite galaxies, so we require a sub-halo mass function.  We use the parameterisation introduced by  \cite{Giocoli2010}: 
 \begin{equation}
\frac{dN(m, M, z)}{dm} = (1+z)^{1/2}\,A_{M}\,M\,m^{\alpha}\exp{\left[-\beta\left(\frac{m}{M}\right)^{3}\right]}\;,
\label{submassf}
\end{equation}
with the parameters $A_{M} = 9.33 \times 10^{-4}$, $\alpha = -1.9$ and $\beta = 12.2715$. Recently, \cite{Dooley2014} have shown that this sub-halo mass function does not strongly depend on the choice of the cosmological parameters.
$N(M,z)$, or the HOD,  is simply the integral of the sub-halo mass function of those galaxies above an observable mass or luminosity threshold. 

The assumption of this sub-halo mass function is that the number of substructures per host halo mass is universal; more massive haloes host proportionately more satellite galaxies.  However, there still remains mass dependence in the exponential cut-off; more massive sub-haloes only exist in more massive haloes.  This latter fact implies a weak size correlation between satellites and their central galaxy hosts, which strengthens if the less massive satellites are not observed. 
 
\subsubsection{Radial profile} 
In addition to knowing how many sub-haloes there are, we need to know how they are distributed around the centre of the halo.   
We assume a Navarro-Frenk-White (NFW) profile \citep{NFW1996} both for the distribution of mass in the halo and for the probability of finding any given  sub-halo at a particular distance from the centre of the halo.   In principle, the sub-halo probability distribution may depend on the sub-halo mass, $m$, and be significantly different from the halo mass distribution.   
  
The NFW profile is given by \citep{NFW1996} 
\begin{equation}
\rho_{\rm{NFW}}(x|M) = \frac{\rho_{\rm s}}{x/r_{\rm s}(1+x/r_{\rm s})^{2}},
\label{nfw}
\end{equation}
where $x$ is the distance from the centre of the halo and $r_{\rm s}$ is the scale radius of the halo. Its concentration is defined as $c=R_{200}/r_{\rm s}$ where $R_{200}$ is the virial radius of the halo.
The virial radius is defined as the radius enclosing an overdensity equal to $200\rho_{\rm cr}$, where $\rho_{\rm{cr, 0}} = 2.7\times 10^{11} M_{\odot}\,{\rm h}^{2}\,{\rm Mpc}^{-3}$ is the critical density of the universe at redshift $z=0$. In particular, including the redshift dependence of the critical density, we have:
\begin{equation}
R_{200}(z) = \left(\frac{3\,M}{4\pi\rho_{\rm cr}(z)\Delta}\right)^{1/3}\;,
\end{equation}
where $\Delta = 200$ is the redshift independent overdensity parameter and $\rho_{\rm cr}(z) = \rho_{\rm{cr,0}}E(z)^{2}$, where $E(z)$ is the expansion rate given by:
\begin{equation}
E(z) =  \frac{H(z)}{H_0}= \sqrt{\Omega_{\rm{m,0}}(1+z)^{3} +\Omega_{\rm{K}}(1+z)^2+\Omega_{\Lambda,0}}\;.
\end{equation}
The normalisation $\rho_{\rm s}$ is given by:
\begin{equation}
\rho_{\rm s}=\frac{M}{4\pi r_{\rm s}^{3}}\left[\ln(1+c) - \frac{c}{1+c}\right]\;,
\end{equation}
and we use the following model for the concentration from \cite{Oguri2011}:
\begin{equation}
c(z) = c_{0}(1+z)^{-0.71}\left(\frac{M}{M_{c_{0}}}\right)^{-0.086}\;,
\end{equation}
where $c_{0}= 7.26$ and $M_{c_{0}}= 10^{12}M_{\odot}\,{\rm h}^{-1}$.  This implicitly assumes that the concentration is a deterministic function of the halo mass, with no scatter. 

We can convert from a matter distribution to a sub-halo probability distribution by simply dividing by the total halo mass, $u(x|M) = \rho_{\rm{NFW}}(x|M)/M$. 
Below we work in Fourier space for calculating power spectra, where it is useful to have the Fourier transform of the normalised density profile given in eq. (\ref{nfw}):
\begin{equation}
u(k|M) = \int_{0}^{R} dx \frac{4\pi x^{2}}{M}\frac{\sin(kx)}{kx}\rho_{\rm{NFW}}(x|M)\;,
\label{ft-nfw}
\end{equation}
In principle we should also specify the radial profiles and mass-concentration relations for the sub-haloes, as in \cite{Giocoli2010}; however, below we assume a simple relation of the satellite radii to the sub-halo mass, so the sub-halo profiles are not required.   

\subsubsection{Large scale halo distribution} 
The final element in the halo model description is to specify the large scale distribution of haloes; this is usually done through specifying two-point (and higher) moments to match the expected linear or weakly non-linear behaviour.   Here we focus on matching the two-point moments by assuming a simple deterministic bias that is mass dependent.  

In the halo model, the two-point correlation function can be written 
\begin{equation}
\xi(\bmath{x}) = \xi_{\rm{1h}}(\bmath{x}) + \xi_{\rm{2h}}(\bmath{x})\;,
\end{equation}
where the first term describes the contribution from each halo whereas the second term gives the contribution on large scales from halo correlations. The mass function and probability density profiles are needed to evaluate both terms, but the two-halo term also requires the halo correlation function $\xi_{\rm{hh}}(\bmath{x}|M_1, M_2)  = b(M_{1})b(M_{2})\xi_{\rm{lin}}{(\bmath{x})}$ where $\xi_{\rm{lin}}(\bmath{x})$ is the linear mass correlation function and $b(M, z)$ is the bias parameter.   
We use the bias model (consistent with the mass function) from \cite{ShethTormen1999}:
\begin{equation}
b(M) = 1+ \frac{q\nu - 1}{\delta_{c}} + \frac{2p}{\delta_{c}(1+(q\nu)^{p})}\;,
\end{equation}
where $p, q$ and $\nu$ are defined as above. 

This approximation is justified because on large scales the density correlation function has to follow the linear correlation function.
There is an explicit constraint on $b(M)$, as pointed out by \cite{Seljak2000}, because on large scale the amplitude of the two-halo term of the mass-weighted density power spectrum has to match the amplitude of the linear power spectrum. This gives a constraint for the halo model bias:
\begin{equation}
\int_{0}^{\infty} dM n(M) b(M)\frac{M}{\bar{\rho}} = 1\;,
\label{constrbias}
\end{equation}
so that, on the very largest scales where the mass profile of the haloes is unimportant, the mass distribution matches linear theory.   

\subsection{Size-mass relation}
As we are interested in the sizes of galaxies and how they are correlated, we must have a process for relating the observed size of a galaxy to the halo model.  For this, we use the size-virial radius relation found by \cite{Kravtsov2013} where abundance matching was used to relate simulated halo masses to the properties of observed galaxies; by this means he found a linear relation between the virial radius $R_{200}$ of the haloes and the radius enclosing half of the galaxy mass $r_{1/2}$:
\begin{equation}
r_{1/2} = 0.015\,R_{200}\;.
\label{halfmassrad}
\end{equation}
\cite{Kravtsov2013} finds that this relation holds over eight orders of magnitude in stellar mass and for all morphological types. 

This relation is consistent with the model developed by \cite{Mo1998} in which galaxy disc sizes are determined by the angular momentum they acquire during the collapse. As also stated in \cite{Kravtsov2013}, it is remarkable that the relation given in eq. (\ref{halfmassrad}) seems to be valid even for early-type galaxies, showing that angular momentum is extremely important in the process of galaxy formation. Additionally,  $r_{1/2}$ can be related to the effective radius of a galaxy $R_{\rm e}$, which is the radius enclosing half of the light of the galaxy, through $r_{1/2} = 1.34\,R_{\rm e}$ \citep{Kravtsov2013}.
In the following we identify $r_{1/2}$ with $r(m)$ in order to keep the notation concise.

\subsection{Mass threshold}
In order to translate the halo model into observable quantities, we need to model the galaxy selection effects.  For simplicity, we will assume that we 
have a survey complete to some intrinsic luminosity threshold.  Assuming the luminosity directly relates to stellar mass,  we require a relationship between halo mass and galactic stellar mass for selecting a minimum halo mass for our calculations. We use the relation given by \cite{Guo2010}:
\begin{equation}
\frac{M_{*}}{M} = C \times \left[ \left(\frac{M}{M_{0}}\right)^{-a} + \left(\frac{M}{M_{0}}\right)^{b} \right]^{d}\;.
\label{Mgal}
\end{equation}
where $C=0.129$, $M_{0}=10^{11.4} M_{\odot}$, $a=0.926$, $b=0.261$ and $d=2.440$ and $M$ is the mass of the host halo. This relation is obtained assuming a one-to-one correspondence between sub-haloes and galaxies by using abundance matching, the hypothesis that the cumulative halo mass function is equal to the cumulative galaxy mass function. 

By means of eq. (\ref{Mgal}) we choose minimum masses for both sub-haloes and haloes equal to $m_{\rm{min}}= M_{\rm{min}} = 10^{11} M_{\odot}\,{\rm h}^{-1}$ that corresponds to a minimum galaxy mass equal to  $2\times10^{9} M_{\odot}\,{\rm h}^{-1}$. Setting this limit for the minimum halo mass is also in agreement with the Halo Occupation Distribution (HOD) model analysed in \cite{Kravtsov2004}.

\section{Translating to observations}
\label{sizecorr}
Given the halo model assumptions, we can work out its implications for observables.  We first look at background quantities before moving on to the two-point quantities of primary interest.   Here, following the approach given in \cite{Sheth2005b}, we build up from the simplest halo model quantities to the size-weighted galaxy distribution and how it impacts the magnification estimator defined above.  

\subsection{Halo density}
We begin with the discrete distribution of the haloes, which is described by 
\begin{equation}
n_{\rm h}(\bmath{x}) =  \int_{0}^{\infty} dM\, \sum_{i} \delta_{\rm D}(M-M_{i}) \delta_{\rm D}^{(3)}(\bmath{x}-\bmath{x}_{i})
\end{equation}
where we have integrated over the possible halo masses.   The sum within the integral has expectation given by the mass function defined above, 
\begin{equation}
\left\langle\,\sum_{i} \delta_{\rm D}(M-M_{i}) \delta_{\rm D}^{(3)}(\bmath{x}-\bmath{x}_{i})\,\right\rangle\, = n(M) = \frac{dN_{\rm h}}{dMdV}\;.
\end{equation}
The total halo density is given by the integral, $ \bar{n}_{\rm h} = \int_{0}^{\infty} dM\, n(M).$

\subsection{Halo matter density}
\label{halodens}
If we assume that the mass distribution is dominated by that associated with the haloes (ignoring that in sub-haloes), 
the dark matter density field is given by:
\begin{equation}
\rho(\bmath{x}) = \sum_{i} \rho_{\rm{NFW}} (\bmath{x}-\bmath{x}_{i}, M_{i}) = \sum_{i} M_{i} u(\bmath{x}-\bmath{x}_{i}, M_{i})\;,
\label{discrete-rho}
\end{equation}
where the sum is over the haloes and $u(\bmath{x}, M)$ is the density profile normalised to the halo mass.
We can obtain a continuous density field from the discrete one given in eq. (\ref{discrete-rho}) by introducing Dirac delta functions:
\begin{equation}\begin{split}
\rho(\bmath{x}) = \int_{0}^{\infty} dM\, M  \int d^{3}x'   \sum_{i} & \delta_{\rm D}(M-M_{i})  \delta_{\rm D}^{(3)}(\bmath{x'}-\bmath{x}_{i})  \\&\times \, u(|\bmath{x}-\bmath{x'}|, M)\;.
\end{split} 
\end{equation}
Taking the ensemble average we obtain the mean matter density: 
\begin{equation}
\bar{\rho} \equiv \langle \rho(\bmath{x}) \rangle = \int_{0}^{\infty} dM\, n(M)\, M\;,
\label{constrdensity}
\end{equation}
where we used the fact that $\int d^{3}x\,u(|\bmath{x}-\bmath{x}_{i}|, M) = 1$ (since the function $u$ is normalised for each halo). 

\subsection{Galaxy density}
\label{galdens}
In the halo model, it is assumed that the galaxy density is composed of two terms, the central galaxies positioned at the halo centre and satellite galaxies distributed around the halo centre.  Analogously to the halo density defined above, we can write the galaxy density as  
 \begin{equation}
n_{\rm g}(\bmath{x}) =  \int_{0}^{\infty} dM\, \sum_{i} \delta_{\rm D}(M-M_{i}) \sum_{j} \delta_{\rm D}^{(3)}(\bmath{x}-\bmath{x}_{i}-\bmath{x}_{j}), 
\end{equation}
where the $\sum_{j}$ is over the central and possible satellite galaxies and $ \bmath{x}_{j}$ represents their position relative to the halo centre; $ \bmath{x}_{j}=0$ for the central galaxy, while for the satellite galaxies, these positions are described by the satellite probability profile. 

The average number of satellites for a halo of a given mass is $\langle N_{\rm sat}| M \rangle$ which is related to the halo occupation distribution; it is an integral of the sub-halo mass function defined above: 
 \begin{equation} 
\langle N_{\rm sat}| M \rangle =  \int_{m_{\rm{min}}}^{M}dm \, \frac{dN(m,M)}{dm}\;,
\label{nsat}
\end{equation}
and the HOD has one more than this to account for the central galaxy. Again, we assume that substructures inside a halo follow a spatial distribution $u_{\rm d}(|\bmath{x}-\bmath{x_{\rm c}}|, M)$ (which we assume to be of the form given in Eq. \ref{nfw}) depending on the halo mass and where $ \bmath{x_{\rm c}}$ are the coordinates of the centre of the halo.
After averaging over the sub-halo ensembles, the galaxy density can be written as 
\begin{equation}
n_{\rm g}(\bmath{x}) =  \sum_{i} \delta_{\rm D}^{(3)}(\bmath{x}-\bmath{x}_{i}) + \langle N_{\rm sat}| M_i \rangle  u_{\rm d}(\bmath{x}-\bmath{x}_{i}| M_{i})\;,
\end{equation}
which can again be written as 
\begin{equation}\begin{split}
n_{\rm g}(\bmath{x}) = & \int_{0}^{\infty}  dM\,   \int d^{3}x'   \sum_{i}  \delta_{\rm D}(M-M_{i})  \delta_{\rm D}^{(3)}(\bmath{x'}-\bmath{x}_{i})  \\&\times \, 
\left[ \delta_{\rm D}^{(3)}(\bmath{x}-\bmath{x'}) + \langle N_{\rm sat}| M \rangle u_{\rm d}(|\bmath{x}-\bmath{x'}|, M)\right]\;.
\end{split} 
\end{equation}
After averaging over the positions of the haloes, we find 
\begin{equation}
\bar{n}_{\rm g} = \int_{0}^{\infty}  dM\,  n(M) (1 + \langle N_{\rm sat}| M \rangle).
 \end{equation}
This could alternatively be written as 
\begin{equation}\begin{split}
 \bar{n}_{\rm g} = \int_{M_{\rm{min}}}^{+\infty} dM\, & n(M)\int_{m_{\rm{min}}}^{M} dm \\
  &\times \left(\delta_{\rm D}(m-M) + \frac{dN(m,M)}{dm}\right) \;,
  \end{split}
 \end{equation}
 where we have introduced minimum halo and galaxy masses which will arise in realistic observations and assumed the central galaxy mass is comparable to that of the halo itself.  Though a fraction of the total halo mass will reside in the sub-haloes, we do not expect this to greatly impact the central galaxy size.   

\subsection{The galaxy size field} 
Our assumption is that the observed half-radius is related to the sub-halo mass, as described above. We weight the galaxy density defined in Sec. \ref{galdens} by the radius, and normalise by the total galaxy density to define a galaxy size field as 
 \begin{equation}
r(\bmath{x}) =  \bar{n}_{\rm g}^{-1} \int_{0}^{\infty} dM\, \sum_{i} \delta_{\rm D}(M-M_{i}) \sum_{j} \delta_{\rm D}^{(3)}(\bmath{x}-\bmath{x}_{i}-\bmath{x}_{j}) r(m_j) 
\label{sizewg}
\end{equation}
where $r(m_j)$ is the radius associated with the mass of the central galaxy or satellite galaxies.
For the central galaxy, we should formally base its radius on the residual mass, that is, subtracting the integrated mass in sub-haloes from the total halo mass.  
However, even for the smallest haloes in our model, the total mass in the sub-haloes only accounts for around 10\% of the halo mass, so this correction makes a small change in the inferred central radius. 
We checked that our correlation results are not affected by making this correction and for simplicity we adopt the radius based on the total mass.   

The galaxy size field given by eq.(\ref{sizewg}) can be averaged over the sub-halo ensembles to find 
 \begin{equation}\begin{split}
r(\bmath{x}) & = \bar{n}_{\rm{g}}^{-1}\sum_{i} [r(M_i) \delta_{\rm{D}}^{(3)}(\bmath{x}-\bmath{x}_{i}) \\ 
& +\bar{r}_{\rm{sat}}(M_i) \langle N_{\rm{sat}}| M_i \rangle  u_{\rm{d}}(\bmath{x}-\bmath{x}_{i}| M_{i})] \\
& = \bar{n}_{\rm{g}}^{-1} \int_{0}^{\infty} dM\,   \int d^{3}x'   \sum_{i}  \delta_{\rm{D}}(M-M_{i})\delta_{\rm{D}}^{(3)}(\bmath{x'}-\bmath{x}_{i}) \\
& \times [ r(M) \delta_{\rm{D}}^{(3)}(\bmath{x}-\bmath{x'}) + \bar{r}_{\rm{sat}}(M) \langle N_{\rm{sat}}| M \rangle u_{\rm{d}}(|\bmath{x}-\bmath{x'}|, M)]\;.
\end{split}
\end{equation}
where $\bar{r}_{\rm{sat}}(M)  \equiv \int_{m_{\rm{min}}}^{M}\frac{dN(m,M)}{dm} r(m) dm /\langle N_{\rm{sat}}|M\rangle $ is the average satellite radius for satellites in a halo of mass $M$.  

\begin{figure}
\centering
\includegraphics[scale=0.6]{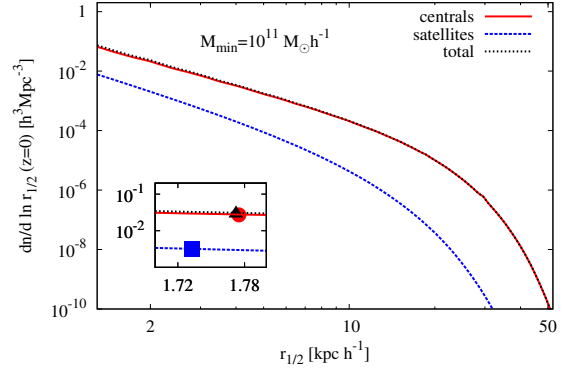}
\caption{Distribution of object for given half-mass radius for both haloes and sub-haloes as well as for the total population at redshift $z=0$ and for minimum halo mass $M_{\rm min} = 10^{11} M_{\odot}\,{\rm h}^{-1}$. Symbols indicates the mean half-mass radii for each population (square for satellites, circle for centrals, triangle for the total population). In the main plot the number densities are shown for the entire range of half-mass radii; for a better view of the mean values the inset plot represents the number densities in the range between $r_{1/2} = 1.7\,\rm{kpc}\,{\rm h}^{-1}$ and $r_{1/2} = 1.8\,\rm{kpc}\,{\rm h}^{-1}$ .}
\label{distrmass}
\end{figure}

With this, it is straight forward to 
derive the distribution of radii and derive the average galaxy size:
\begin{equation}\begin{split}
 \bar{r} = \bar{n}_{\rm{g}}^{-1}&\int_{M_{\rm{min}}}^{+\infty} dM\,n(M)\int_{m_{\rm{min}}}^{M} dm  \\
  &\times \left(\delta_{\rm{D}}(m-M) + \frac{dN(m,M)}{dm}\right) r(m)\;.
  \end{split}
 \end{equation}
In Fig.\ref{distrmass} radii distributions, calculated by means of the size-mass relation given by eq. (\ref{halfmassrad}) found by \cite{Kravtsov2013} combined with the halo mass function, for centrals, satellites and total galaxy population are shown and the mean values for half-mass radius for each type of structures are indicated at redshift $z=0$.

\subsection{The local estimator field} 

The magnification estimator, defined as
\begin{equation}
\hat{\kappa} = \ln{\frac{\lambda_{\rm O}}{\bar{\lambda}}} - \left\langle \ln{\frac{\lambda_{\rm O}}{\bar{\lambda}}} \right\rangle\;,
\end{equation}
acts on the observed angular sizes of galaxies, potentially combining galaxies over a range of redshifts.      
It is possible however to consider a local definition of the estimator field that when summed over redshift becomes the two-dimensional projected estimator.   
 
The intrinsic contribution to the magnification estimator arises because the observed size depends on the true galaxy size.  These are related through the angular diameter distance, and for objects at a given redshift $\lambda_{\rm I} = r(z)/D_A(z) $, so that, 
\begin{equation}
\ln{\frac{\lambda_{\rm I}}{\bar{\lambda}}} = \ln{\frac{r(z)}{\bar{r}}} + \ln{\frac{\bar{r}}{D_{\rm A}(z) \bar{\lambda}}}.
\end{equation}
For objects at a given redshift, their observed size field and true size field are related by a constant term, which cancels when considering the fluctuation field.   Their fluctuations are identical,
\begin{equation}
\ln{\frac{\lambda_{\rm I}}{\bar{\lambda}}} - \left\langle \ln{\frac{\lambda_{\rm I}}{\bar{\lambda}}} \right\rangle_{z}\ = \ln{\frac{r(z)}{\bar{r}}} - \left\langle \ln{\frac{r}{\bar{r}}} \right\rangle_{z}\;.
\end{equation} 
Thus, the intrinsic contribution is effectively 
\begin{equation}
\kappa_{\rm I} (z) = \ln{\frac{r(z)}{\bar{r}}} - \left\langle \ln{\frac{r}{\bar{r}}} \right\rangle_{z}\;.  
\label{eqn:kI}
\end{equation}
Note that in both cases, dividing by the mean radius (or angular size) makes the argument of the logarithm dimensionless, but any scale would be equivalent, as the divisors cancel when subtracting the field average.  It is the clustering of relative sizes which contributes to the magnification estimator. 

This work is primarily concerned with statistics of the angular sizes of galaxies, projected over a broad redshift distribution.  
Statistics related to the true physical sizes of galaxies potentially would be biased by individual photometric redshift errors, and so it would be essential to treat these carefully in any 3-D or tomographic analysis of the physical size correlations.  

A given realisation of halo and sub-halo positions results in a estimator-weighted density field as 
 \begin{equation}\begin{split}
\kappa_{\rm I}(\bmath{x}) =  n_{\rm g}^{-1} \int_{0}^{\infty} dM\, \sum_{i} & \delta_{\rm D}(M-M_{i}) \\
 &\times \sum_{j}\delta_{\rm D}^{(3)}(\bmath{x}-\bmath{x}_{i}-\bmath{x}_{j}) \kappa_{\rm I}(m_j).   
\end{split}
\end{equation}
%We define the estimator for satellites as follows:
%\begin{equation}\begin{split}
%\kappa_{I, s}(\bmath{x}) = & \bar{n}^{-1}\int_{M_{\rm{min}}}^{+\infty} dM\,n(M) \int_{m_{\rm{min}}}^{M} dm \frac{dN(m, M)}{dm}  \\ 
%&\times \kappa_{I}(m) u_{d}(|\bmath{x}-\bmath{x}_{c}|, M)\;,
%\end{split}
%\label{massfield}
%\end{equation}
where:
\begin{equation}
\kappa_{\rm I}(m,z) =  \ln{\left(\frac{r(m)}{\bar{r}}\right)} - \left\langle \ln{\left(\frac{r}{\bar{r}}\right)}\right\rangle_{z}\;.
\label{hatkappa}
\end{equation}
By definition, the expectation of this estimator is zero, $\langle \kappa_I \rangle = 0; $
the expectation value of the log-size field at a given redshift is 
\begin{equation}\begin{split}
\left\langle \ln{\left(\frac{r}{\bar{r}}\right)}\right\rangle_{z} &= \bar{n}^{-1}_{\rm g} \int_{M_{\rm{min}}}^{\infty} dM\, n(M)\int_{m_{\rm{min}}}^{M} dm \\ 
&\times \left(\delta_{\rm D}(m-M) + \frac{dN(m, M)}{dm} \right) \ln{\left(\frac{r(m)}{\bar{r}}\right)}\;.
 \end{split}
\end{equation}

\section{Two-point statistics} 
\label{2pointStat}
Our focus here is to understand the implications of size correlations on two-point statistics, and in particular in comparing how the power spectrum of the magnification estimator relates to that of the true magnification once size correlations are included.  
Thus, we must calculate the power spectrum of the intrinsic size correlations and their cross correlation with the true magnification.  

 As discussed above, in the halo model two-point correlations receive contributions from pairs of galaxies inhabiting the same halo and from where they inhabit two different haloes.  
The same holds for the power spectrum:
\begin{equation}
P(k) = P_{\rm{1h}}(k) + P_{\rm{2h}}(k)\;.
\end{equation}
It is straightforward to calculate the power spectrum of the matter density fluctuation $\delta \rho/\bar{\rho}$  using the halo model formalism developed above  \citep{Scherrer1991}: 
\begin{equation}\begin{split}
P_{\rm{1h}}(k) &= \int_{0}^{\infty} dM n(M) \left(\frac{M}{\bar{\rho}}\right)^{2} u^{2}(k, M)\;, \\
P_{\rm{2h}}(k) &= \bar{b}_{\rho}^{2} P^{\rm{lin}}(k)\;,
\end{split}
\end{equation}
where
\begin{equation}
\bar{b}_{\rho} = \int_{0}^{\infty} dM n(M) b(M)\frac{M}{\bar{\rho}} u(k, M)\;.
\label{brho}
\end{equation}
%Here we are interested in calculating the two-point statistics \citep{Scherrer1991}:
%\begin{equation}\begin{split}
%\langle &\sum\nolimits_{i}  \delta_{D}(M_{1}-M_{i}) \delta_{D}^{(3)}(\bmath{x}_{1}-\bmath{x}_{i}) \sum\nolimits_{j} \delta_{D}(M_{2}-M_{j}) \times \\ & \times\delta_{D}^{(3)}(\bmath{x}_{2}-\bmath{x}_{j})\rangle = n(M_{1})\delta_{D}(M_{1}-M_{2})\delta_{D}^{(3)}(\bmath{x}_{1}-\bmath{x}_{2}) + \\ &+ n(M_{1})n(M_{2})\xi_{hh}(\bmath{x}_{1}-\bmath{x}_{2}, M_{1}, M_{2})\;.
%\end{split}
%\label{dots}
%\end{equation}
%where $\xi_{hh}$ is the halo correlation function discussed above. 

The number density fluctuation is similar, but accounts for the central and satellite galaxy contributions separately.   The one-halo term includes terms from the central-satellite and the satellite-satellite pairs within the same halo: 
\begin{equation}\begin{split}
P_{\rm{1h}}(k) = \bar{n}_{\rm g}^{-2} \int_{0}^{\infty} & dM  n(M) (\langle N_{\rm sat}|M\rangle u(k, M)  \\ & + \langle N_{\rm sat}(N_{\rm sat}-1)|M\rangle u^2(k, M)\  ) \;.
\end{split}\end{equation}
The two-halo term has three contributions, including central-central, central-satellite and satellite-satellite terms: 
\begin{equation}\begin{split}
P_{\rm{2h}}(k) = \bar{b}_{n}^{2} P^{\rm{lin}}(k),
\end{split}\end{equation}
where 
\begin{equation}
\bar{b}_n = \bar{n}_{\rm g}^{-1} \int_{0}^{\infty} dM n(M) b(M) \left(1+ \langle N_{\rm sat}|M\rangle u(k, M)\right) \;.
\label{bn}
\end{equation}

\subsection{Magnification estimator power spectrum}

In this subsection we present our model for the correlation between log-size of galaxies. In the one-halo terms, we only include the cross-correlations between different galaxies, so there is no central-central contribution.  

%For galaxies located at the centre of each halo our log-size field is:
%\begin{equation}
%\kappa_{I, c}(\bmath{x}) =  \bar{n}^{-1}\int_{M_{\rm{min}}}^{+\infty} dM\,n(M) \kappa_{I}(M) \delta^{\rm{lin}}(\bmath{x})\;,
%\end{equation}
%where $\delta^{\rm{lin}}(\bmath{x})$ is the linear matter density fluctuation field.

\subsubsection{One-halo terms}

Applying the halo model formalism, we obtain the following power spectra for the auto-correlation:
\begin{equation}\begin{split}
P^{\rm{1h-sat}}_{\kappa_{\rm I}}(k) &= \bar{n}_{\rm g}^{-2} \int_{M_{\rm{min}}}^{\infty} dM n(M)\\
&\times\left[\int_{m_{\rm{min}}}^{M} dm \frac{dN(m, M)}{dm}\kappa_{\rm I}(m) u_{\rm d}(k, M)\right]^{2} \end{split}
\end{equation}
We also have contribution from central-satellite correlation terms:
\begin{equation}\begin{split}
P^{\rm{1h-cs}}_{\kappa_{\rm I}}(k) &= \frac{2}{\bar{n}_{\rm g}^2} \int_{M_{\rm{min}}}^{\infty} dM n(M)\,\kappa_{\rm I}(M) \\  &\times\int_{m_{\rm{min}}}^{M} dm \frac{dN(m, M)}{dm} \kappa_{\rm I}(m) u_{\rm d}(k, M) 
\end{split}
\end{equation}

\subsubsection{Two-halo terms}

Applying the halo model formalism, we obtain the following power spectra for the auto-correlation:
\begin{equation}\begin{split}
P^{\rm{2h}}_{\kappa_{\rm I}}(k) = (\bar{b}_{\kappa_{\rm{I, c}}}+ \bar{b}_{\kappa_{\rm{I, s}}})^{2}P^{\rm{lin}}(k)\;, 
\end{split}
\end{equation}
where:
\begin{equation}
\bar{b}_{\kappa_{\rm{I, c}}} = \bar{n}_{\rm g}^{-1} \int_{M_{\rm{min}}}^{\infty} dM\,n(M)\,b(M)\,\kappa_{\rm I}(M)\;
\label{barbm}
\end{equation}
and 
\begin{equation}\begin{split}
\bar{b}_{\kappa_{\rm{I, s}}} = \bar{n}_{\rm g}^{-1}\int_{M_{\rm{min}}}^{\infty} &dM n(M)b(M)\,\int_{m_{\rm{min}}}^{M} dm \frac{dN(m, M)}{dm} \\ &\times \kappa_{\rm I}(m) u_{\rm d}(k,M)\;.
\end{split}
\end{equation}

\begin{figure}
\centering
\includegraphics[scale=0.6]{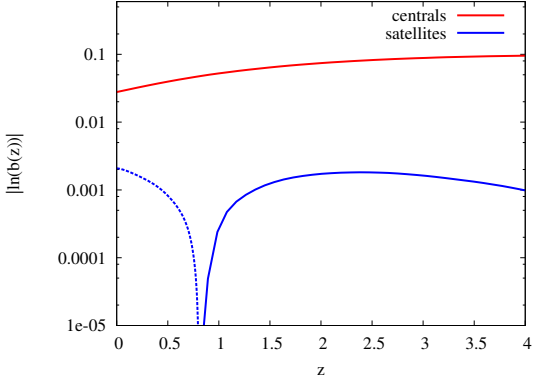}
\caption{The bias factors arising from central and satellite galaxies, as a function of redshift, for a survey with fixed mass threshold.  Satellites are more numerous, but have a lower average size.  Those on the lower threshold are most numerous and dominate at low redshift; as these are below the average size, the satellite bias becomes negative. }
\label{fig:bias}
\end{figure}

These biases are perhaps the most important result of our model, as the two-halo terms dominate on the scales where lensing is most easily interpreted. 
In Fig.~\ref{fig:bias} we show how the central and satellite biases evolve as a function of redshift. 
The central bias ranges from 0.1 at high redshifts, down to a few times $10^{-2}$ at low redshifts, while the satellite bias is considerably smaller ($\sim 10^{-3}$), becoming negative at low redshifts. 

As the sample will be dominated by central galaxies at this mass threshold, it is worth trying to understand its amplitude better in the limit where there are only central galaxies.  Recall the definition of the intrinsic kappa field is the log of the radius minus its average (see eq. (\ref{eqn:kI})). 
Examining the expression for the central bias, we see that it is effectively a weighted average of 
$\kappa_{\rm I}$, where the number density weight is modified by a bias function, $b(M).$   Were it not for this bias factor, this integral is the usual density averaging, meaning that the two terms in $\kappa_{\rm I}$ would exactly cancel by definition.   

If $b(M)$ were constant, independent of the mass, the central bias would also be zero.  The central bias thus depends on how $b(M)$ changes as a function of mass.  In particular, since the bias increases for larger mass haloes, where the radii are larger than average, this implies $\bar{b}_{\kappa_{\rm{I, c}}}$ is positive.  Its magnitude depends on how fast $b(M)$ increases over the mass range that dominates the estimator, $M_{\rm min} < M < 10^{14} M_{\odot}\,{\rm h}^{-1}$.

The picture is somewhat more complex when the satellite population becomes more important.   The satellite distribution is weighted somewhat to lower mass galaxies (Fig. 1), so the mean of the log radius becomes smaller.  This fact tends to increase the central bias.  Meanwhile, the weighting towards lower mass tends to cancel the increase in $b(M)$, reducing the amplitude of $\bar{b}_{\kappa_{\rm{I, s}}}$.  For the lowest redshifts, the up-weighting of the low masses is enough to make $\bar{b}_{\kappa_{\rm{I, s}}}$ negative.

\begin{figure*}
\centering
\includegraphics[scale=0.6]{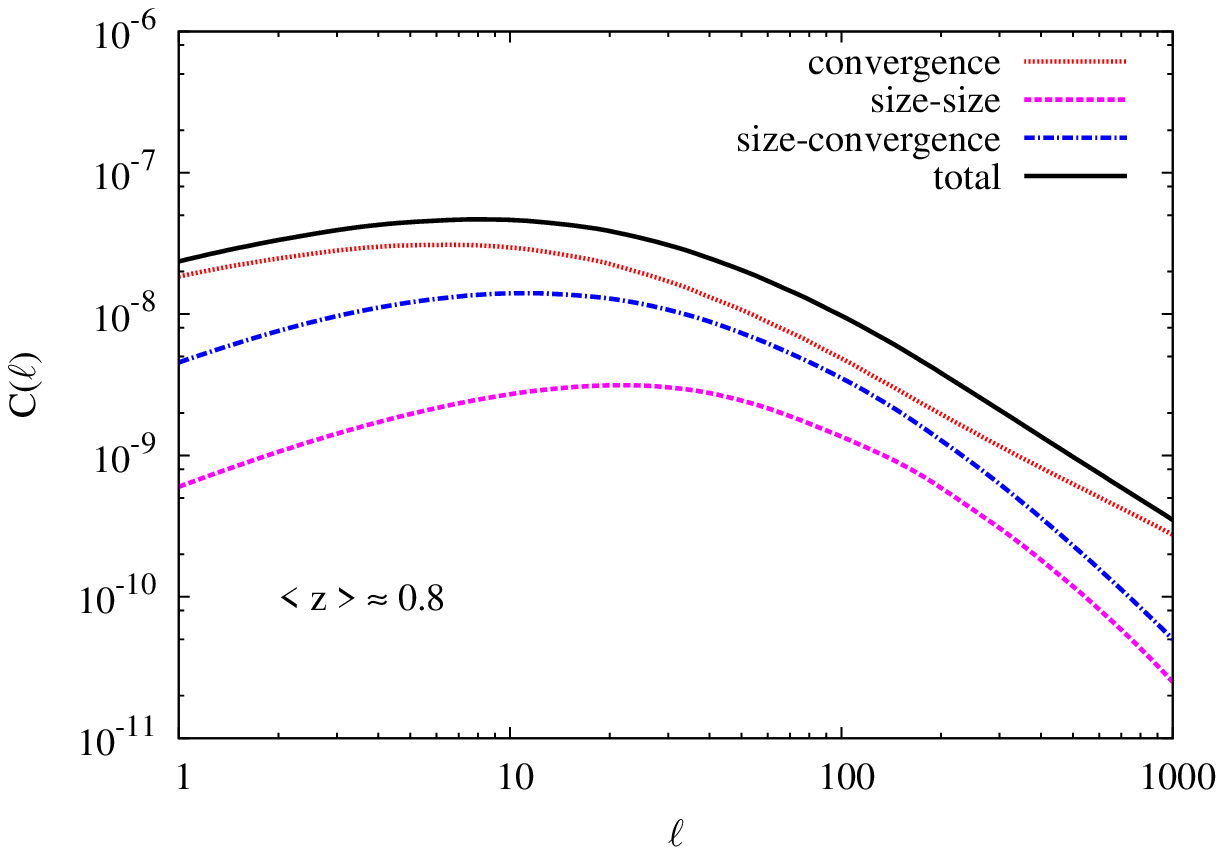}
\includegraphics[scale=0.6]{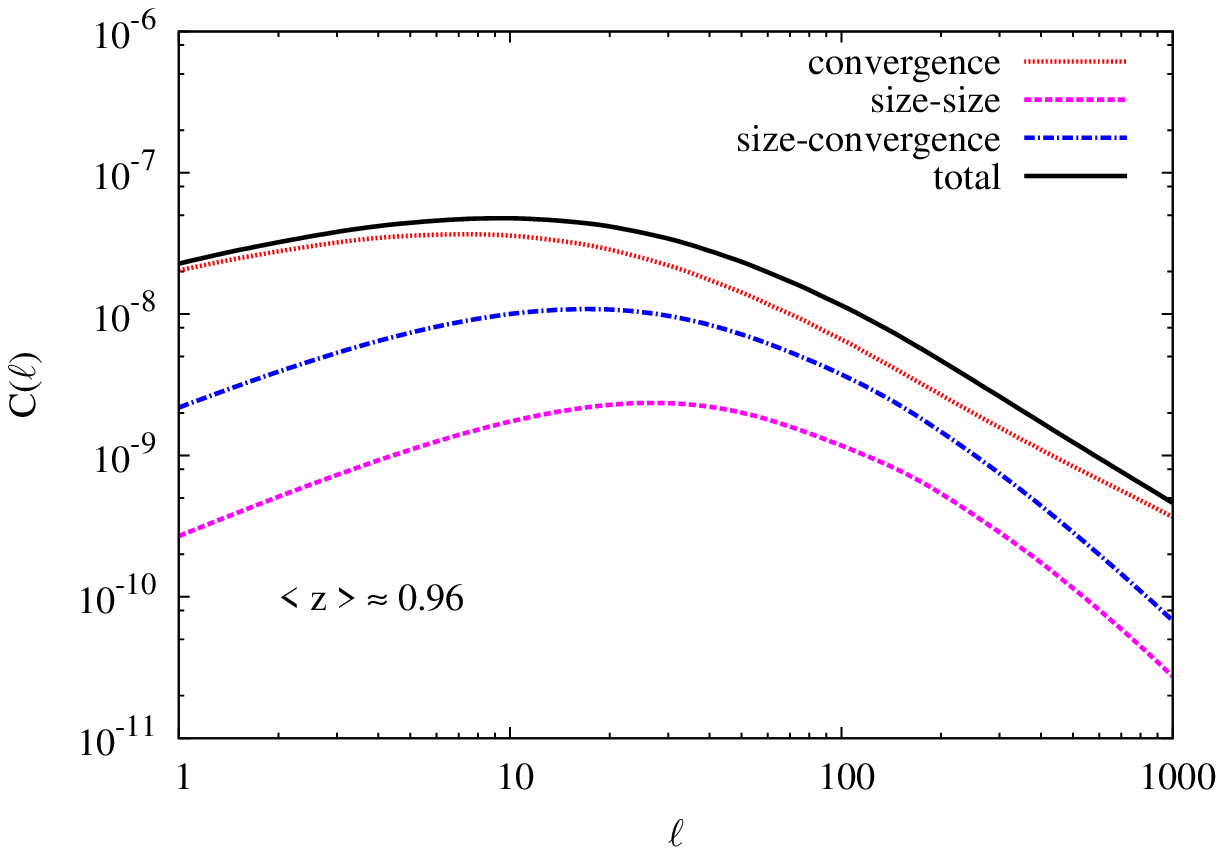}
\caption{Power spectra for two different redshift distributions, CFHTLenS-like with mean redshift $\langle z \rangle \simeq 0.8$ and Euclid-like with $\langle z \rangle \simeq 0.96$.}
\label{Fig:powspectra}
\end{figure*}

\subsection{Density-size cross power spectra}
For the cross-correlation density-size we obtain for both central and satellites:
\begin{equation}\begin{split}
P_{\rho\kappa_{I}}^{\rm{1h-sat}} (k) &= \bar{\rho}^{-1}\bar{n}_{\rm g}^{-1} \int_{0}^{\infty} dM n(M) M  \\
&\times\int_{m_{\rm{min}}}^{M} dm\frac{dN(m, M)}{dm} \kappa_{\rm I}(m)\,u(k, M)\, u_{\rm d}(k, M) \\
P_{\rho\kappa_{\rm I}}^{\rm{2h} }(k) &= \bar{b}_{\rho}(\bar{b}_{\kappa_{\rm{I, c}}} +\bar{b}_{\kappa_{\rm{I, s}}}) P^{\rm{lin}}(k) 
\end{split}
\end{equation}
where $\bar{b}_{\rho}$ is given in eq. (\ref{brho}) (using the constraints given in eq. (\ref{constrdensity}) and eq. (\ref{constrbias})) and the other bias factors are given above. 

In this work, we are assuming all of the lensing mass is associated with the haloes, and ignore mass associated with sub-clumps.  On large scales, this should be a good approximation, but potentially it fails to take into account further correlations between size and density on scales within haloes.  It would be straight forward to extend this work to include this effect in the halo model. 

\subsection{Angular power spectra}

In order to compare intrinsic size correlations with weak lensing convergence power spectra we have to integrate the projected size correlations over the redshift distribution:
\begin{equation}
\kappa_{\rm I}(\theta) = \int d\chi\,n(\chi)\,\kappa_{\rm I}(\chi\theta, \chi)\;,
\end{equation}
where $n(\chi)$ is the redshift distribution described in section \ref{lensing-sec}. Again we assume Limber's approximation and the total convergence power spectrum can be written as:
\begin{equation}
C_{\hat{\kappa}}(\ell) = C_{\kappa}(\ell) + 2C_{\kappa\kappa_{\rm I}}(\ell) + C_{\kappa_{\rm I}}(\ell)\;.
\label{pklambda}
\end{equation}
The lensing term is given by eq. (\ref{convergence}) and the intrinsic terms in eq. (\ref{pklambda}) are calculated as follows:
\begin{equation}\begin{split}
C_{\kappa\kappa_{\rm I}}(\ell) &= \int_0^{\chi_{\rm{hor}}} d\chi \, 
\frac{q(\chi)n(\chi)}{\chi^2} \, P_{\rho\kappa_{\rm I}} \left( \frac{\ell}{\chi},\chi \right) \\
C_{\kappa_{\rm I}}(\ell) &= \int_0^{\chi_{\rm{hor}}} d\chi \, 
\frac{n^{2}(\chi)}{\chi^2} \, P_{\kappa_{\rm I}} \left( \frac{\ell}{\chi},\chi \right)\;,
\end{split}
\label{angularpl}
\end{equation}
where $q(\chi)$ is the lensing weight function defined in section \ref{lensing-sec}.

\section{Results}
\label{results}

\subsection{Model assumptions }

We evaluate our results in the context of a flat $\Lambda$CDM cosmology with parameters consistent with best-fit Planck data \citep{Planck2013-params}; in particular, we assume a total matter density  $\Omega_{\rm{m,0}} = 0.32$, cosmological constant density $\Omega_{\Lambda,0} = 0.68$, baryon density  $\Omega_{\rm{b,0}} = 0.049$ and Hubble constant $H_{0} = 100\,{\rm h}\,{\rm km}\,{\rm s}^{-1}\,{\rm Mpc}^{-1}$, where $h = 0.67$.  In addition, we assume the spectral index of the matter power spectrum is $n_{\rm s} = 0.96$ and it is normalised such that $\sigma_{8} = 0.83$.

We adopt the transfer function given in \cite{EisensteinHu1998} and non-linear evolution of the matter power spectrum (for estimating lensing convergence power spectrum) is calculated with HALOFIT from \cite{Smith2003} recently revised by \cite{Takahashi2012}.  
 
For the redshift distribution of lensed sources, we adopt the commonly used parameterisation, 
\begin{equation}
n(z) \propto z^{a}\exp{\left[-\left(\frac{z}{z_{0}}\right)^{b}\right]}\;.
\label{redsource}
\end{equation}
We consider two different set of parameters for this redshift distribution form; following \cite{Schneider2010}, to simulate a Euclid-like survey we assume $a=2$, $b=1.5$, $z_{0}=0.64$ which gives a mean redshift around $0.96$.  For a CFHTLenS-like survey, we use parameters from \cite{Benjamin2007}: $a=0.836$, $b=3.425$, $z_{0}=1.171$ which give a mean redshift approximately $ z \simeq 0.8$. For the shallow survey we used $a=0.6$, $b=1.5$, $z_{0}=0.55$ in order to obtain a mean redshift around $ z \simeq 0.5$.

\subsection{Comparison of power spectra}

In Fig.~\ref{Fig:powspectra} we show the contributions to the power spectrum of $\hat{\kappa}$ for the CFHTLenS and Euclid-like surveys.  As can be seen, intrinsic size correlations are relevant even for a very deep survey such as Euclid, where their contamination increases from 10\% on the largest scales to being comparable to the convergence on $\ell \sim 100.$  
For the CTHTLenS-like survey, with $\langle z \rangle \simeq 0.8$,  the contamination is even larger, beginning at  $25\%$ of the convergence signal on large scales. 

For these surveys,  the largest intrinsic contribution comes from the cross correlation between the intrinsic sizes and the convergence, while the intrinsic auto-correlation is sub-dominant except at the smallest scales.  On the largest scales, $\kappa$ and $\kappa_{\rm I}$ are strongly correlated as the ratio $\langle \kappa\kappa_{\rm I} \rangle/\sqrt{\langle\kappa\kappa\rangle \langle\kappa_{\rm I}\kappa_{\rm I} \rangle}$ is of order 80\%.  

\begin{figure}
\centering
\includegraphics[scale=0.6]{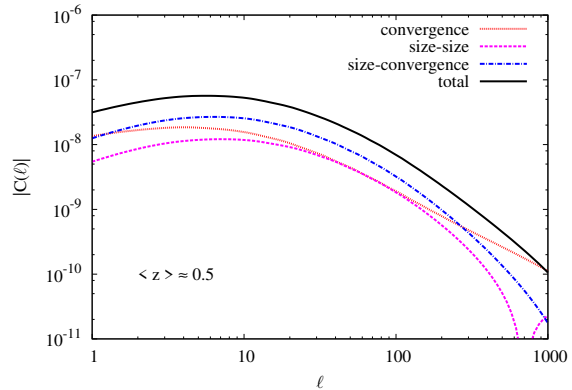}
\caption{The power spectra for a more shallow survey, with mean redshift around $z \sim 0.5$.  Here the intrinsic effects are more significant than for surveys centred at higher redshifts. }
\label{fig:shallow}
\end{figure}

For a shallower redshift distribution, the intrinsic contamination can dominate the signal.  To demonstrate this, in Fig.~\ref{fig:shallow} we show the contributions for a survey with $\langle z \rangle \simeq 0.5$; there we see the intrinsic and convergence spectra are comparable, and significantly correlated.  
With multiple bins, the convergence dominates in high redshift bins, but remains correlated with the intrinsic sizes in lower redshift bins; unlike the convergence, the intrinsic sizes will be relatively uncorrelated between bins.  
% Can you confirm this?   
%(Discussion of narrow bin?  Since the convergence arises at lower redshifts, the cross-correlation is considerably reduced for a narrow bin at high redshifts.  However, the size correlations can introduce non-negligible correlations between convergence observed in high redshift galaxies and the intrinsic sizes of galaxies at lower redshifts. N-by-N plot of correlations?)

\begin{figure*}
\centering
\includegraphics[scale=0.6]{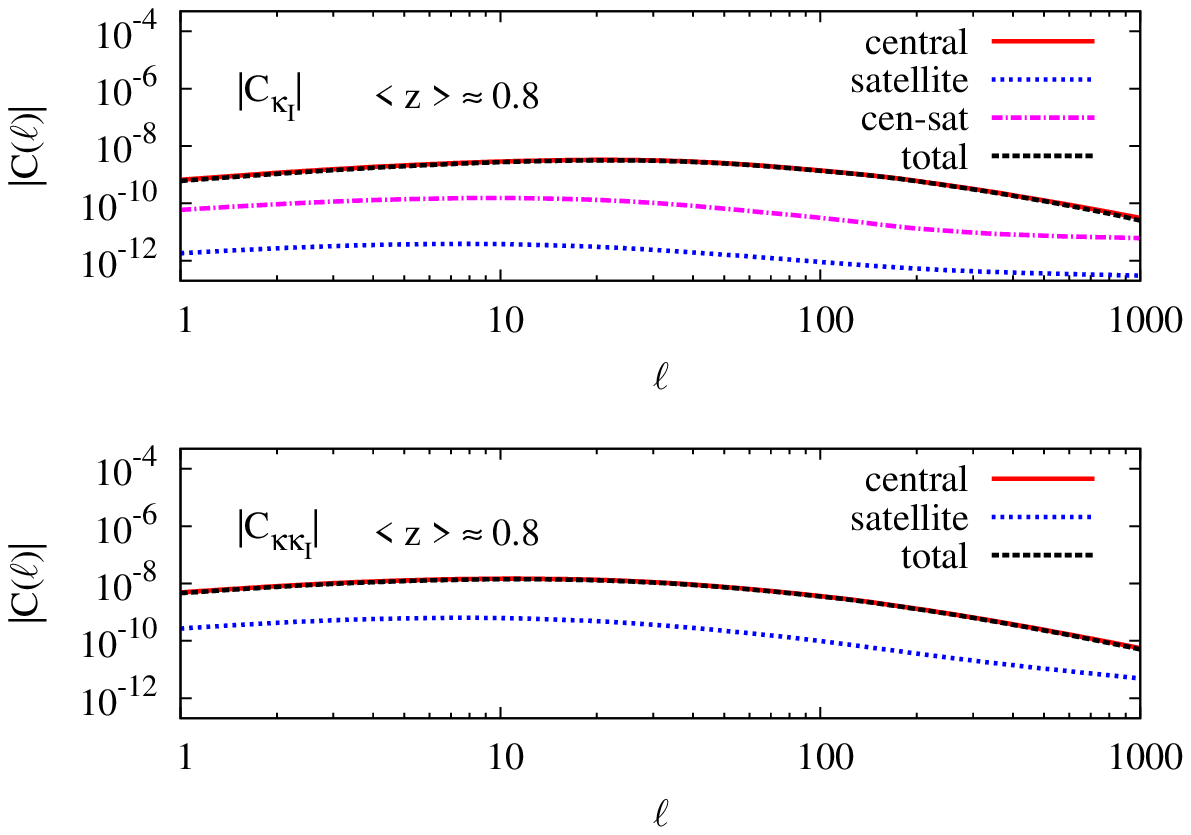}
\includegraphics[scale=0.6]{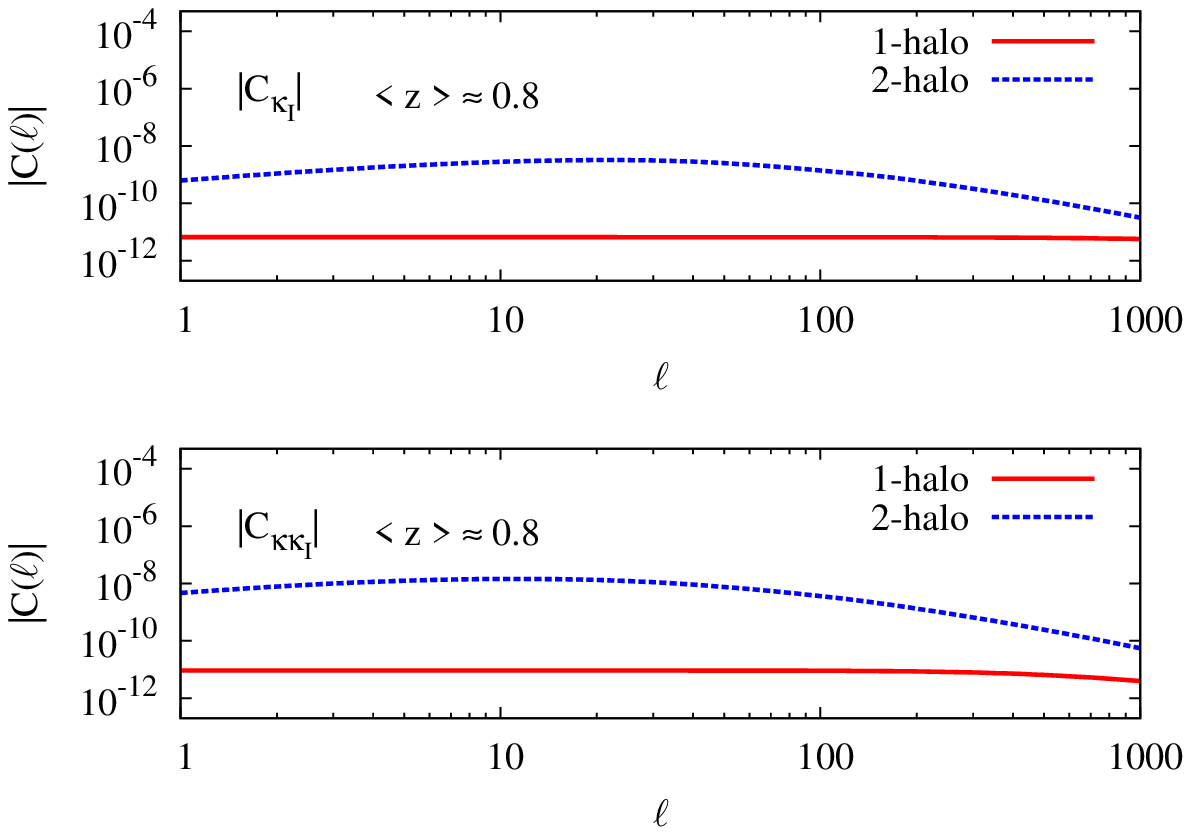}
\caption{Contributions to the intrinsic size (upper plots) and intrinsic size-convergence (lower plots) power spectra for CFHTLenS-like survey ($\langle z \rangle \simeq 0.8$).  We plot the absolute values; the central-satellite contribution for $C_{\kappa_{\rm I}}$  and the satellite contribution for $C_{\kappa\kappa_{\rm I}}$ as well as the one-halo term are negative. On the scales of interest, the correlations are dominated by the two-halo contributions for the central galaxies.}
\label{contribution}
\end{figure*}

In Fig.~\ref{contribution} we show the contribution to the size-size power spectrum and size-convergence power spectrum arising from centrals and satellites, and also how the spectra arise from the one-halo and two-halo terms.  The spectra are dominated by the two-halo contributions on the scales of interest, and on these scales the central galaxy contribution is most significant; this follows from what was seen previously for the central and satellite biases.  On smaller scales, the one-halo term and the contribution from satellites both become more important. 

Recall that at low redshift, the satellite bias becomes negative, because satellite galaxies have sizes generally smaller than the total mean value.  In the size-size correlation, this leads to the satellites being negatively correlated with the central galaxy population.  They also contribute negatively to the size-convergence spectrum, though with an amplitude much smaller than the positive amplitude arising from the central sizes. 

Formally the size-size power spectrum should be positive definite; however here we have omitted correlations of galaxies with themselves.  As a result, on small scales the negative cross-correlation between central and satellite galaxies can actually dominate.  On such scales, probing the typical galaxy sizes, our model is not expected to be physical; on these scales, galaxies will begin to overlap and they would not be observed as distinct.  

%Also the central-satellite contribution to the size-size power spectrum is negative and this can again be due to the fact that satellites are smaller than the total mean size. Contributions from central and satellite populations are shown in Fig. \ref{contribution} for different redshift distributions; the satellite contribution is much more important on small scale and affects in general all scale since it is a negative contribution. In the intrinsic size power spectrum the satellite contribution is positive but we can notice that the central-satellite term has approximately the same amplitude of the central term but with opposite sign and, therefore,  the satellite contribution is dominant.

%The size-size power spectrum is always positive and its importance increases moving to larger and larger $\ell$ where sub-haloes become important. In particular the size-size power spectrum can be as large as $30\%$ for $\ell \sim 50$ and become the dominant signal for $\ell >500$.  Because of these intrinsic correlations, the final estimated lensing convergence signal is over estimated up to $\ell \sim 200$ on average more than $20\%$.
%The intrinsic correlations are smaller at higher redshift whereas the convergence increases its amplitude; this is due to the different weight functions in the projection integrals in eq. (\ref{angularpl}). Therefore the intrinsic contributions are lower for deeper survey but still non negligible as it is shown on right panel in Fig.~\ref{Fig:powspectra} and in Fig.~\ref{contribution}.

\section{Conclusions}
\label{conclusions}

We have presented a simple model for calculating intrinsic correlations for galaxy sizes using halo model formalism.   This is a first calculation and necessarily neglects some effects which could be very relevant.  One important issue that should be factored in is scatter in the mass-radius relation; this could considerably weaken the correlations we see in the sizes.  Galaxy sizes may also be environmental  dependent and affected by baryonic physics in ways that are hard to fold into the simple halo model.  

We also have restricted our analysis to a simple mass threshold in the selection of galaxies.  For a more realistic analysis, one might consider how these effects would impact a flux limited sample, or one with a cut-off in the observed angular size of galaxies.  We plan to continue our study by examining the magnitude of the effect in galaxy surveys like the Sloan Digital Sky Survey, focussing on low redshifts where intrinsic effects should dominate. 

%We find that sub-haloes are extremely important in this model and their contribution may be not negligible when convergence power spectrum is estimated through observed size correlations in the sky. In particular we find an overestimate of the convergence power spectrum on large scale and negative estimate on small scales where the satellite galaxies contribution becomes important. This could be due to the fact that satellite galaxies are the dominant population with respect to the central galaxies and, on average, are smaller giving negative values for our logarithmic estimator. We obtain this to be an important effect on small scales. 

Our preliminary study indicates that, as for measurements of galaxy shapes, it may not be possible to ignore intrinsic correlations when interpreting measurements of galaxy sizes and magnitudes.  These effects, and particularly correlations between convergence and intrinsic properties, are potentially an important systematic for magnification measurements and could significantly bias the resulting cosmological constraints if they are not accounted for.  On the other hand, they represent a new observable that could potentially tell us more about how galaxies form. 

Correlations of galaxy magnitudes are also used to detect magnification and these are similarly expected to be correlated with halo masses; it is worth investigating how magnitudes are correlated with both convergence and galaxy sizes and this is a straight forward extension of the halo model we have developed here.  We plan to pursue this in future work.

\section*{Acknowledgments}

We thank David Bacon, Alan Heavens, Enrique Gaztanaga, Catherine Heymans and Elisabeth Krause for useful conversations.  RC and FP acknowledge support from STFC grant ST/H002774/1.  This work was also supported in part by the National Science Foundation under Grant No. PHYS-1066293 and the hospitality of the Aspen Center for Physics.

\bibliographystyle{mn2e}

\bibliography{IntrinsicSize_v2.bbl}
\bsp

\label{lastpage}

\end{document}